\documentclass{PoS}
\usepackage{epsfig}

\newcommand{\bea}{\begin{eqnarray}}
\newcommand{\eea}{\end{eqnarray}}
\newcommand{\nn}{\nonumber}
\newcommand{\vdir}{v\kern -3.pt\raise 0.15ex\hbox {/}}
\newcommand{\1}{1\kern -4.pt\raise 0.01ex\hbox {1}}

\title{Use and misuse of ChPT in the heavy-light systems~\thanks{Supported in part by the EU Contract No.~MRTN-CT-2006-035482, ``Flavianet".}}

\ShortTitle{Use and misuse of ChPT for the heavy-light systems}

\author{\speaker{Damir Be\'cirevi\'c}
 \\
	Laboratoire de Physique Th\'eorique (B\^at. 210)\\
	Universit\'e Paris-Sud, Centre d'Orsay \\
	F-91405 Orsay-Cedex, France\\
        E-mail: \email{Damir.Becirevic@th.u-psud.fr}}

\author{Svjetlana Fajfer\\
        J. Stefan Institute, Jamova 39, P.O.Box 3000, and \\
        Department of Physics, University of Ljubljana,\\
	Jadranska 19, 1000 Ljubljana, Slovenia \\
        E-mail: \email{Svjetlana.Fajfer@ijs.si}}

\author{Jernej Kamenik\\
        INFN, Laboratori Nazionali di Frascati \\
	I-00044 Frascati, Italy\\
        E-mail: \email{Jernej.Kamenik@lnf.infn.it} }

\abstract{We discuss the range of validity of chiral perturbation theory 
when applied to the systems of heavy-light mesons. Having in mind the recent 
experimental evidence according to which the heavy-light 
scalar and axial states are closer to the ground states
than anticipated, we revisited the prediction for the chiral behavior 
of the $B^0_q-\overline B^0_q$ mixing amplitude and examined the
impact of nearness of the $(1/2)^+$ states. We conclude that the 
standard ChPT expressions 
with $N_F=3$ light flavours are not useful in guiding the extrapolation 
of  hadronic quantities computed on the lattice. Instead those derived in HMChPT 
with $N_F=2$, i.e., including only the pion loops, are still adequate as long as they 
are applied to the pions lighter than $350$~MeV, or the quarks lighter than a third of 
 the physical strange quark mass.
}

\FullConference{The XXV International Symposium on Lattice Field Theory\\
		 July 30-4 August 2007\\
		 Regensburg, Germany}

\begin{document}

\section{Introduction}
In this note we summarise the findings of our 
research presented in ref.~\cite{ourBB}. 
The fact that the lattice QCD community is heavily  
dependent on the formulae derived in chiral perturbation theory (ChPT) 
to extrapolate the directly accessible results for almost any phenomenologically
relevant quantity to the physical --nearly chiral--
limit, requires a clear assessment of the validity of 
ChPT. Only then one can be able to claim 
``a high precision physical result" deduced from the lattice QCD simulations 
combined with ChPT. 
Before touching on the peculiarities related to the heavy-light mesons, 
we will make a short trip to the sector of light mesons because the first 
warnings of how far we should (should not) push ChPT in terms of precision
requirements already show there. 

\subsection{Light mesons}
\noindent
The viability of  chiral expansion in the theory with $N_F=3$ light flavours 
($u$, $d$, and $s$) has been questionable since the very beginning of the theory. 
The main reason is that the strange quark mass is half-a-way between the chiral
limit and $\Lambda_{\rm QCD}$, which might significantly lower the ChPT order parameters 
[${\displaystyle{\lim_{m_{u,d,s}\to 0}}}(f_\pi,\langle \bar q q\rangle)$], with respect to their $N_F=2$ counterparts, 
and thus spoil the perturbative nature of the chiral expansion (a number of situations 
in which the next to leading order term in the chiral expansion is larger than the leading one). 
Of course, the issue can be settled after confronting the experimental data to the ChPT 
formulae, but the quality of the actual $K\pi$-scattering data is still 
not good enough to resolve this issue (a special worry is related to the low energy constants 
$L_4$ and $L_6$). As of now, it is safe to say that the formulae derived in 
ChPT with $N_F=3$ are not sufficiently reliable to seek ${\cal O}(1\ \%
)$ 
precision when extrapolating the lattice data to the physical limit.

\begin{figure}[h]
\begin{center}
\includegraphics[width=8.3cm,clip]{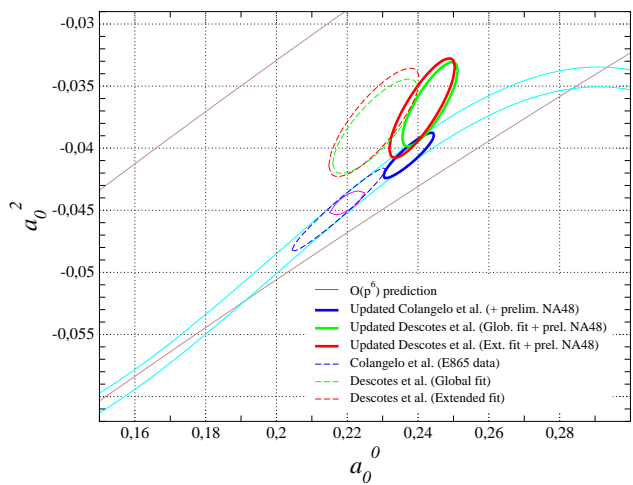}
\caption{\label{fig1}\sl $S$-wave scattering lengths $a_0^I$ (isospin $I=0,2$). 
The dashed (thick) ellipses are obtained by using the old (new) experimental 
$K_{e 4}$-data. The two sets of ellipses refer to two different 
analyses~\cite{Colangelo,Descotes} while the small one, in the centre of
the plot, is the 2-loop ChPT prediction (Courtesy of S.Descotes-Genon). }
\end{center}
\end{figure}

The situation with $N_F=2$ is different simply because that theory has been tested.
Back in 2001, such a test looked like a triumph: the experimentally measured $S$-wave 
scattering lengths of $\pi\pi$-system, emerging from $K_{e 4}$ decay, were 
fully consistent with the NNLO ChPT formula. It was deduced that the 
the Gell-Mann--Oakes--Renner formula (GMOR), $m_\pi^2 = 2 B_0 m_{u,d} + \dots$, 
is  saturated by the leading term to more than $94\%
$~\cite{Colangelo}.  However, that situation changed since the new 
and improved $K_{e 4}$ data, collected in NA48, appeared.
If one repeats exactly the same analysis as that of ref.~\cite{Colangelo} 
with those new data then ``only" about $85\%
$ of the GMOR formula is saturated by the first term, which can be read off from the plot 
shown in fig.~\ref{fig1}.

In spite of the speculations that the electromagnetic and isospin breaking corrections 
might patch up the new conclusion we believe it is fair to summarise that ChPT 
with pions only ($N_F=2$) passed the experimental tests but the level of accuracy is still 
controversial. Instead, the actual situation in the $N_F=3$ case remains unclear.

 \subsection{Heavy-light mesons}
\noindent
ChPT has been extensively applied to describe the dynamics of 
light constituents in the heavy-light mesons. In particular,
heavy-meson ChPT (HMChPT) has been constructed in the static heavy 
quark limit~\cite{casalbuoni}, with only one extra parameter, $g$, the coupling of a 
pseudo-Goldstone boson (PGB) to a doublet of lowest lying 
heavy-light mesons [$j_\ell^P=(1/2)^-$], namely the pseudoscalar ($J^P=0^-$)
and vector ($1^-$) mesons. 
HMChPT with $N_F=2$ and $N_F=3$ light flavours inherit the problems discussed above. 
An extra complication appeared after the experimenters 
reported on the observation of the orbitally excited states [$j_\ell^P=(1/2)^+$], both scalar ($0^+$) 
and axial ($1^+$) ones. The observations are made in the case of mesons with the charmed heavy quark~\cite{experiment}, 
to which the confirmation came from the unquenched lattice study in the static heavy quark limit~\cite{green}. 
In summary the splitting between the excited and ground states is only
\bea
\Delta_{S_s} \equiv m_{D_{0s}^{\ast}}-m_{D_s}=m_{D_{1s}}-m_{D_s^\ast}=350~{\rm MeV},\qquad
\Delta_{S_q}\approx 430(30)~{\rm MeV}\,, 
\eea 
for the strange and the non-strange case respectively. 
Leaving aside the reasons why $\Delta_{S_s}\neq\Delta_{S_q}$, it is clear that 
$\Delta_{S_{s,q}} < \Lambda_\chi, m_\eta, m_K$, and thus the basic assumption that no resonances 
 appear between zero and $\Lambda_\chi$ is simply not correct in HMChPT.~\footnote{The fact that $\Delta_{S_s}<\Delta_{S_q}$ 
is still craving for an explanation. It cannot be reconciled with results from the lattice simulations, 
 nor with the HMChPT considerations~\cite{ourMs}.}

In view of importance of the HMChPT guidance to extrapolating the lattice results to 
reach the phenomenologically relevant physical quantities, 
such as $f_B$, or $B^0-\overline B^0$ mixing parameters, we revised the derivation of these expressions in HMChPT and studied
the impact of $\Delta_{S}$ on the chiral behavior of the $B^0_d-\overline B^0_d$ mixing amplitude. 
The special attention is given to $B^0_d-\overline B^0_d$, rather than $B^0_s-\overline B^0_s$, 
because of its better potential for the new physics search~\cite{bona}.

\section{Computation of $f_B\sqrt{m_B}\to \hat f_q$}

In HMChPT, in the static heavy quark limit ($m_Q\to \infty$), the pseudoscalar decay constant  
is of dimension $3/2$, and schematically we write ${\displaystyle{
\lim_{m_B\to \infty} f_{B_q}\sqrt{m_{B_q}}}}\to \hat f_q$. Its light quark dependence is expected to be 
described by the chiral loops shown in fig.~\ref{fig:g}, and the result reads  
\bea\label{fb1}
&&\hat f_d = \hat f_0 \left[ 
1 - {1+3g^2 \over 4 (4\pi f)^2 }
 \left( 
 3 m_\pi^2\log{m_\pi^2\over \mu^2} +  2 m_K^2\log{m_K^2\over \mu^2} 
  + {1\over 3}  m_\eta^2\log{m_\eta^2\over \mu^2} \right)
 +  {\rm c.t.} \right]\,,
\eea
where $\hat f_0$ is the heavy-light meson decay constant in the chiral limit, $g$ is the coupling of the 
heavy-light mesons doublet to a PGB (also in the chiral limit) are the new parameters 
are the two constants coming from the weak current and from the lagrangian.    
``c.t." stands for the local counterterms the $\mu$-dependence of which cancels 
against the one present in the chiral logarithms.   \EPSFIGURE[r]{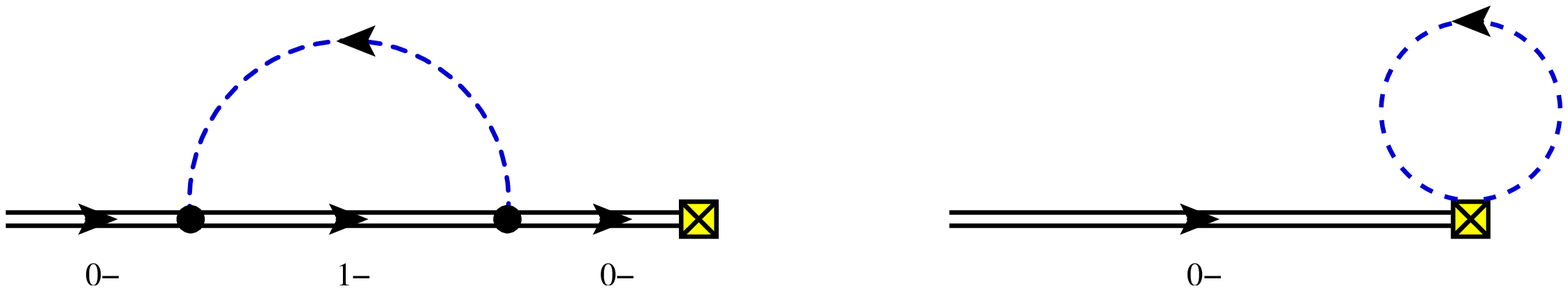,width=7.8cm}{\label{fig:g}
The chiral loop corrections to $\hat f_d$: the box denotes the weak current vertex, double line 
is the heavy-light meson and the dashed line is the PGB propagator.}

A tacit assumption in deriving the above formula is that there is a clear 
separation by which the chiral logarithms describe the long distance dynamics whereas 
the local counterterms encode the information on short distance physics. 
The lattice results obtained at larger $m_q > m_d$ are expected to provide 
a viable method to fix those counterterms, i.e., to compute the 
associated low energy constants. 
An equivalent statement is that HMChPT can be used to guide the extrapolation 
of $\hat f_q$ computed for $m_q > m_d$  to the physical limit. 
  \EPSFIGURE[l]{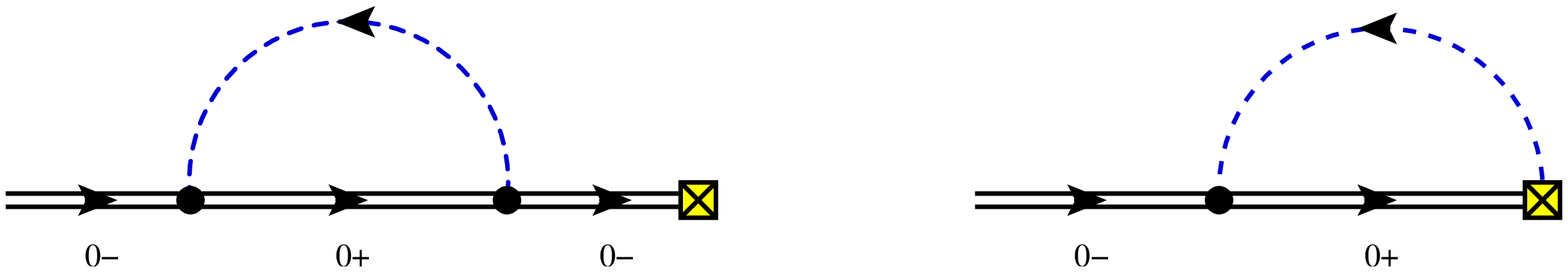,width=7.8cm}{\label{fig:h}
Inclusion of the scalars in the chiral corrections to $\hat f_d$. Notice that $m_{0^+}- m_{0^-}=\Delta_S < \Lambda_\chi$. }
Specifying the separation scale means distinguishing the particles which can propagate 
in the chiral loops from those which cannot. In eq.~(\ref{fb1}) that scale is evidently assumed to be $\Lambda_\chi \gtrsim 
m_\eta$.~\footnote{ The propagation of $1^-$ heavy-light state does not give any contribution because we take it to be degenerate in mass 
with the pseudoscalar meson,  which holds true in the static heavy quark limit.}
But the fact that $\Delta_{S_{s,q}} \lesssim  m_{K,\eta}$, and thus also $\Delta_{S_{s,q}}<\Lambda_\chi$, means 
that one must include the propagation of the scalars in the loops as well. The lagrangian 
and the weak current that include the effect of
$(1/2)^+$ states  are given in
ref.~\cite{ourBB}. The only two new diagrams that contribute to $\hat f_d$ are those 
shown in fig.~\ref{fig:h} and there are only two new couplings
that appear in the new --extended to include scalar-- expression,  
\bea
&&\hat f_{q} = \hat f_{0} \left\{ 
1 + \sum_i{ t^i_{qa}t^{i\dagger}_{aq} \over 2 (4\pi f)^2}  \biggl[ 
3 g^2 \lim_{x\to 0} {d\over dx}[xJ_1(m_i^2,x)] - I_1(m_i^2) 
-h^2 \biggl(
J_1(m_i^2,\Delta_S) + J_2(m_i^2,\Delta_S) \biggr.\biggr.\right.
\nn\\
&&\hspace*{.6cm} \left.\left.\biggl.+ \Delta_S {d\over d\Delta_S}\left(
 J_1(m_i^2,\Delta_S) + J_2(m_i^2,\Delta_S) \right)
\biggr)
- 2 h{\hat f^+_0 \over \hat f_0 }\left(
I_1(m_i^2) + I_2(m_i^2,\Delta_S) \right) 
\right] + {\rm c.t.}\right\}\,,
\eea
are $h$, the coupling of the PGB to one $(1/2)^-$ and one $(1/2)^+$ heavy-light meson, and $\hat f_0^+$, the
weak decay constant of the orbitally excited heavy-light meson in the chiral limit. 
If one takes the limit $m_i< \Delta_S$, which only applies to pions, there is an amusing 
automatic separation between the $\Delta_S$-dependent terms from the independent ones. The result is that 
the HMChPT with the pion
loops ``survive" whereas the $K$- and $\eta$-loops are drowned in a whole lot of new terms 
of which it is significant the presence of those proportional to $\Delta_S^2\log(\Delta_S^2/\mu^2)$, thus 
competitive in size with the $K$- and $\eta$-contributions. This definitely restricts the applicability  
of the HMChPT formulae to the light quark masses corresponding to pions lighter than $\Delta_S$.
That can be converted to a condition that the light quark masses to which the HMChPT formulae 
apply should be lighter than a third of the physical strange
quark mass, i.e.,  $m_q < m_s^{\rm phys.}/3$, and only to such data the HMChPT with $N_F=2$ flavours can be used, or
\bea
\hat f_{q} = 
\hat f_0 
\left[ 
1-  {1+3g^2\over 2 (4\pi f)^2} 
{3\over 2} m_\pi^2 \log{m_\pi^2\over \mu^2} 
+ c_f(\mu) m_\pi^2 \right]\,,
\eea 
with $c_f(\mu)$ being the combination of low energy constants that multiply $m_\pi^2$ and which, together with 
$g$ and $\hat f_0$, should be fixed by fitting the lattice data to this expression. 
Otherwise, i.e., if we do not restrict to $m_q < m_s^{\rm phys.}/3$, the number of parameters in the expression 
rises from $3$ to $13$.
\section{Bag Parameters}
The Standard Model (SM) bag parameter in the static heavy quark limit is defined via $\langle \bar B^0_q \vert  \widetilde O(\nu) \vert  B^0_q
\rangle =(8/3) \hat f_q(\nu)^2  \widetilde B_{q}(\nu)$, where the $\Delta F=2$ operator 
$\widetilde O= (\bar b_v\gamma_\mu^L q)(\bar b_v\gamma_\mu^L q)\equiv\gamma_\mu^L\otimes \gamma^\mu_L$.
Assuming its UV-physics ($\nu$-scale dependence) is being taken care of, say in HQET perturbation theory, 
one can study its bosonised form in HMChPT,
\bea
\widetilde  O &=& \sum_X \beta_{1X} {\rm Tr} \left[ (\xi \overline H^Q)_q \gamma_{\mu}
(1-\gamma_5) X \right] \times  {\rm Tr} \left[ (\xi H^{\bar Q})_q \gamma^{\mu} (1-\gamma_5) X \right] + {\rm ct.}\,,\nn \eea
where $X\in\{ 1, \gamma_5, \gamma_{\nu},
\gamma_{\nu}\gamma_5, \sigma_{\nu\rho}\}$, and the field  $H(v)$ describes the $(1/2)^-$-doublet, consisting 
of the pseudoscalar and vector heavy-light mesons.
Once the scalar/axial fields are introduced in HMChPT the expressions become much more complicated 
and essentially useless for a meaningful numerical study. In ref.~\cite{ourBB} we showed that for $m_q < m_s^{\rm phys.}/3$, 
we again can separate the
contribution of the pion loop corrections and lump everything else into the finite counterterms. Before giving the explicit
expressions, let us show on a specific example how that separation occurs.~\footnote{This practice is pretty standard in 
soft collinear effective theory.} Consider a typical (dimensionally regularised) integral that appears 
in these calculations and expand around the decoupling limit of the positive parity states: 
\bea\label{limits2}
&&- 2 (4\pi)^2 v_\mu v_\nu \times
 i \mu^\epsilon  \int {d^{4-\epsilon} p
\over (2\pi)^{4-\epsilon} }{p^\mu p^\nu\over (p^2-m_\pi^2) [\Delta_S^2 - (vp)^2] }\nn\\
&&= - {2 (4\pi^2)\over \Delta_S^2} v_\mu v_\nu \left[ 
 i \mu^\epsilon  \int {d^{4-\epsilon} p
\over (2\pi)^{4-\epsilon} }{p^\mu p^\nu\over p^2-m_\pi^2  } + {\cal O}(1/\Delta_S^2)
\right] \longrightarrow -{m_\pi^4\over 2\Delta_S^2}\log{m_\pi^2\over \mu^2} + \dots  
\eea
where the dots stand for terms of higher order in $m_\pi^2/\Delta_S^2$.  This expansion separates the $N_F=2$ chiral loops from
the rest, diagram by diagram, as demonstrated in ref.~\cite{ourBB}. At the end we arrive at the useful formulas
\bea \label{eq00}
\widetilde B_{q} \hat f_q^2 &=& \widetilde B_{0}\hat f_0^2
 \left[ 
1 - {3 g^2 +2 \over  (4\pi f)^2} m_\pi^2\log{m_\pi^2\over \mu^2} + c_{{\cal O}_1}(\mu)m_\pi^2\right]\,,\nn \\
 \widetilde B_{q} & = & \widetilde B_0 
 \left[ 
1 - {1-3 g^2  \over 2 (4\pi f)^2} m_\pi^2\log{m_\pi^2\over \mu^2} + c_{{B}}(\mu)m_\pi^2\right]\,.
\eea
Note that there is one new counterterm coefficient, $c_{{B}}(\mu)$, and $\widetilde B_{0}={\displaystyle{\lim_{m_q\to 0}
\widetilde B_{q}}}$, both of which should be fixed from lattice data collected with the light quark $m_q < m_s^{\rm phys.}/3$.

\section{Supersymmetric $B^0-\overline B^0$ mixing amplitude}
In SUSY not only $W$-boson propagates in the loop, and thus not only 
the left-left operator survives at low energies. In the static 
heavy quark limit there are in fact $4$ operators, the matrix 
elements of which can contribute to the 
$B^0-\overline B^0$ mixing amplitude. In HMChPT that number 
further reduces to $3$ because the two operators  
differ only by the gluon exchange which cannot alter
the chiral logarithms~\cite{withgio}. The remaining operators 
are $\widetilde  O_{2,4}=\1_L \otimes \1_{L,R}$, where 
$\1_{L/R}=\bar b_v (1\mp \gamma_5) q$. The bosonised forms of these operators were
first correctly figured out in ref.~\cite{lin}, which we then confirmed by
deriving them in a somewhat different way~\cite{ourBB}. The resulting 
expressions for the bag parameters, defined as $\langle \bar B^0_q \vert 
\widetilde O_{2,4} \vert  B^0_q \rangle =(b_{2,4}/3) \hat f_q^2  
\widetilde B_{2,4}$, with $b_2=-5$, $b_4=7$, are of course different from the SM bag-parameter. 
In the theory with  $N_F=2$ light quarks ($u$, $d$) we have
\bea \label{eq22}
\widetilde B_{2,4q} \hat f_q^2 &=&  \widetilde B_{2,4}^{\rm Tree} \hat f_0^2
 \left[ 
1 - {3g^2(3 -Y) +3\pm 1 \over 2 (4\pi f)^2} m_\pi^2\log{m_\pi^2\over \mu^2} 
+ c_{{\cal O}_{2,4}}(\mu)m_\pi^2\right] \,,\cr
\Rightarrow \, \widetilde B_{2,4q}   &=&  \widetilde B_{2,4}^{\rm Tree}  \left[ 
 1 + {3 g^2 Y \mp 1 \over 2 (4\pi f)^2} m_\pi^2\log{m_\pi^2\over \mu^2} 
 + c_{{B}_{2,4}}(\mu)m_\pi^2\right]\,,
 \eea
which structurally differs from eq.~(\ref{eq00}) in that the coefficient multiplying 
the logarithmic contribution here involves a yet another 
low-energy constant, $Y$, and which is also to be extracted from the fit with the lattice data. 
In our paper~\cite{ourBB} we went through the same steps as above: we presented the expressions 
obtained in the theory with $N_F=3$, then included the effects of the scalars in the loops and 
showed that a decoupling of the pion piece with respect to the kaon, eta and the contribution of
excited heavy-light mesons occurs in this case too. Therefore the useful formulas are those written 
in eq.~(\ref{eq22}).

\section{Conclusion}

ChPT is nowadays accepted as an effective theory of QCD at very low energies. 
However, it is a theory solely based on the spontaneous chiral symmetry breaking pattern, 
$SU(N_F)_L\otimes SU(N_F)_R\to SU(N_F)_V$, and tells us nothing about confinement. 
Its elementary objects are PGB's, and not quarks and gluons like in QCD. Therefore 
an appropriate matching of ChPT to QCD amounts to solving the confinement problem in QCD, as well as 
that of the spontaneous chiral symmetry breaking at the more fundamental level.  
In other words, the matching between ChPT and QCD --as of now-- is unclear. 
This is to be contrasted to the case of heavy quark effective theory (HQET) where such a matching 
at any given order in the $1/m_Q$-expansion can be made and is in fact systematically 
improvable, order-by-order in $\alpha_S(m_Q)$.

Both HQET (expansion in $1/m_Q^n$) and ChPT (expansion in $p_\pi^{2n}$) share the same worry, i.e., 
 how good is their convergence in realistic situations in which one retains only one or two 
subleading terms in the expansion.~\footnote{It is perhaps worth mentioning that unlike in ChPT, where expansion is in 
$p_\pi^{2n}$, in HMChPT it goes like $p_\pi^{n}$.} 
In ChPT with $N_F=3$ that issue is still a subject of controversies, while in the $N_F=2$ case 
the experimental tests have been made and the results are quite encouraging although it 
is still unclear how to interpret this test in terms of accuracy.

From the point of view of lattice QCD practitioners, the chiral behavior predicted by an 
effective theory for the quantities that are of high phenomenological interest, such as 
$f_B$, $B^0-\overline B^0$ mixing amplitude, or the $B\to \pi$ form factors, is a very important 
guideline when extrapolating the results collected at unphysical light quark masses down to 
the physical $d/u$-quark mass. The hope is that a result 
of such an extrapolation has smaller systematic errors. 

Spurred by the recent experimental evidence indicating that the heavy-light excited states belonging 
to the $(1/2)^+$ doublet are much lighter than expected, we revisited the predictions 
based on the use of HMChPT with $N_F=3$ light flavours and in the static heavy quark limit. 
We showed that in practical applications only the formulas involving the pion loops, i.e. 
the theory with $N_F=2$, should be used. Otherwise the number of low energy constants to be fixed
from the lattice data becomes prohibitively large and the contributions due to the presence of the near 
heavy-light excitations are comparable in size to the ones that are due to kaon- and/or $\eta$-loops.
We showed on explicit examples how this decoupling occurs and provided the expressions for 
$f_B$, the standard and SUSY $B^0-\overline B^0$ mixing amplitude, as well as for the couplings $g$, $h$~\cite{jernej}, 
the Isgur-Wise function~\cite{svjetlana}, and the scalar meson decay constant~\cite{ourBB}.


\end{document}